\documentclass[modern]{aastex63Rev}
\usepackage{amssymb, amsmath}
\usepackage{graphicx}
\NewPageAfterKeywords
\begin{document}

\title{Constraints on the Binarity of the WN3/O3 Class of Wolf-Rayet Stars}
\altaffiliation{This paper includes data gathered with the 6.5 meter Magellan Telescopes located at Las Campanas Observatory, Chile.}

\author[0000-0001-6563-7828]{Philip Massey}
\affiliation{Lowell Observatory, 1400 W Mars Hill Road, Flagstaff, AZ 86001, USA}
\affiliation{Department of Astronomy and Planetary Science, Northern Arizona University, Flagstaff, AZ, 86011-6010, USA}
\email{phil.massey@lowell.edu}

\author[0000-0002-5787-138X]{Kathryn F. Neugent}
\altaffiliation{NASA Hubble Fellow}
\affiliation{Center for Astrophysics, Harvard \& Smithsonian, 60 Garden St., Cambridge, MA 02138, USA}
\affiliation{Lowell Observatory, 1400 W Mars Hill Road, Flagstaff, AZ 86001, USA}

\author[0000-0003-2535-3091]{Nidia I. Morrell}
\affiliation{Las Campanas Observatory, Carnegie Observatories, Casilla 601, La Serena, Chile}

\begin{abstract}

The WN3/O3 Wolf-Rayet (WR) stars were discovered as part of our survey for WRs in the Magellanic Clouds. The WN3/O3s show the emission lines of a high-excitation WN star and the absorption lines of a hot O-type star, but our prior work has shown that the absorption spectrum is intrinsic to the WR star.  Their place in the evolution of massive stars remains unclear.   Here we investigate the possibility that they are the products of binary evolution.  Although these are not WN3+O3~V binaries,  they could still harbor unseen companions.  To address this possibility, we have conducted a multi-year radial velocity study of six of the nine known WN3/O3s.  Our study finds no evidence of statistically significant radial velocity variations, and allows us to set stringent upper limits on the mass of any hypothetical companion star: for probable orbital inclinations,  any companion with a period less than 100 days must have a mass $<$2$M_\odot$.  For periods less than 10 days, any companion would have to have a mass $<$1$M_\odot$.  We argue that scenarios where any such companion is a compact object are unlikely. The absorption lines indicate a normal projected rotational velocity, making it unlikely that these stars evolved with the aid of a companion star that has since merged. The modest rotation also suggests that these stars are not the result of homogenous evolution.   Thus it is likely that these stars are a normal but short-lived stage in the evolution of massive stars.

\end{abstract}

\section{Introduction}

Wolf-Rayet stars are the evolved descendants of massive O-type stars. Their spectra are dominated by broad, strong emission lines formed in dense stellar winds.  WN-type WRs predominately show lines of nitrogen and helium, the products of the CNO-cycle of hydrogen-burning, while WC-type WRs show carbon and oxygen, the products of helium-burning.  WN-type WRs show little or no hydrogen; the WCs show none.  Although we understand that WRs have been stripped of their outer layers, revealing the products of nuclear burning, we do not know the relative importance of stellar winds and binary interactions in this process.  

In order to better understand the formation mechanism for these massive stars, we conducted a survey to identify a complete sample of WRs in the Magellanic Clouds \citep{MasseyMCWRI,MasseyMCWRII,MasseyMCWRIII,FinalCensus}.  As part of this work, we discovered a new class of these objects, dubbed the WN3/O3 stars \citep{MasseyMCWRI,FinalCensus,NeugentWN3O3s}, further adding to the mysteries of massive star evolution.  In this paper we report on a multi-year project to determine the binarity of the WN3/O3s, placing limits on the masses of any companions, in order to better understand how they formed.

As summarized in \citet{NeugentWN3O3s} and \citet{FinalCensus}, the WN3/O3s show an optical emission-line spectrum
typical of a high-excitation, nitrogen-rich WN3 star, along with an absorption spectrum typical of an O3~V star.  We were immediately able to exclude the possibility
that these were WN3+O3~V binaries as their absolute magnitudes were $M_V=-2$ to $-3$, in accord with what is expected for
a WN3 star alone (see, e.g., \citealt{1986A&A...160..185B,vanderHuchtII}) and much fainter than that of an O3~V ($M_V=-5$ to $-6$; see, e.g., \citealt{1988NASSP.497.....C,MasseyTeff}).  In all, nine such stars were discovered in the Large Magellanic Cloud (LMC).  UV spectra of three
of these, obtained and described by \citet{NeugentWN3O3s} using the Hubble Space Telescope, further demonstrated that the absorption was not coming from a normal O-type star, as there was no sign of the C\,{\sc iv} $\lambda 1550$ resonance wind line which would have otherwise been the strongest UV feature.  \citet{NeugentWN3O3s} conducted a comprehensive spectral analysis, demonstrating that the emission and absorption likely arose from a single source for each star, and that a binary explanation was not necessary to explain the spectral features.  The analysis showed that physically these stars have high effective temperatures (100,000$\pm$5,000~K), a bit hotter than the typical 90,000~K of most other WN3-4s.  Their bolometric luminosities ($\log L/L_\odot=5.6\pm0.3$) are consistent with most LMC WN3-WN4 stars \citep{PotsLMC}.  Their surface abundances are consistent with CNO equilibrium, as are most other WNs, with the only peculiarity that they are still relatively hydrogen-rich, with a He/H number ratio of 1$\pm$0.2, rather than the
usual $>10$.  The other difference of note is that the mass-loss rates are very low for WRs, with $\log{\dot{M}}\sim-5.9\pm0.2$ for all nine stars, rather than the typical $-5.0$~dex for WNs of similar luminosities (see Figure 19 in \citealt{NeugentWN3O3s}).

We describe the WN3/O3s as a new class of WRs, as there are no previously known WRs with similar characteristics.  Absorption lines in the spectra of WRs have usually proven to be the
signatures of OB-type companions, although there are exceptions.  For instance, the  ``slash" stars (``O2-3/WN5-6") show both WR-like emission as well as O-type absorption lines.  The most luminous and massive stars in R136, NGC~3603, and M33's NGC~604 are all such objects; these stars are so close to their Eddington limits that their winds are optically thick, as first
suggested by  \citet{deKoterR136} and \citet{MH98}.  {\it Thus the O2-3/WN5-6 show absorption lines because of their high luminosities, which is not the case for the WN3/O3 stars.}   In the SMC, most
of the WNs show absorption features \citep{2001ApJ...550..713M,MasseyWRSMC}, but most have been shown to be binaries \citep{2003MNRAS.338..360F,2016A&A...591A..22S, 2015A&A...581A..21H}. However,  modeling by \citet{2015A&A...581A..21H} of archival optical data suggests that in AB1 and AB12 the absorption may be intrinsic.  Additional work with better data is underway to determine if these stars are indeed low-metallicity analogs of the LMC's WN3/O3s, but we note here that AB1 and 12 have $M_V$'s of $-4.6$ and $-4.0$, respectively \citep{MasseyWRSMC}, implying a visual luminosity 3-8$\times$ greater than that of the LMC WN3/O3s.  Finally, we note that the Galactic 
 ``WN3+abs" star HD~9974 has never been shown to be a binary.
No radial velocity variations were found by \citet{1981ApJ...244..173M}, who argued that the absorption was intrinsic to the WR.  Modeling by \citet{2004MNRAS.353..153M} showed that the emission and absorption likely arose in the same object, and suggested it was the result of homogenous evolution, in which stars evolve fully (or mostly) mixed due to high initial rotation speeds. 

Where WN3/O3s fall in the evolution of massive stars is unclear \citep{NeugentWN3O3s}.  Their spatial distribution is the same as that of other WRs in the LMC \citep{NeugentWN3O3s,FinalCensus}, and thus we can infer that they formed out of the same metallicity as other LMC WNs. There are 142 ``true" WRs known in the LMC (excluding the O2-3If/WN5-6 stars; see \citealt{FinalCensus} and \citealt{RSGWRs}), so the WN3/O3s represent 6\% of the LMC WR population, or 8\% of the LMC's WNs.  This suggests that these are not so rare that they require special circumstances for their formation.  One possibility is that these WN3/O3s are in a relatively short-lived transitional phase between O-type stars and hydrogen-poor WNs, and that they will develop denser winds as they 
lose their hydrogen envelopes.   Another possibility is that these stars are the results of homogenous evolution, as had been suggested for HD~9974.  However \citet{NeugentWN3O3s} argue against this possibility, noting that the projected rotational velocities of all of the WN3/O3s are a modest 120-150 km s$^{-1}$ rather than the high rotational speeds one requires for homogenous evolution.  While one or two WN3/O3s might be viewed at an unfavorable inclination, it is highly unlikely that is true for the entire sample.  Although mass loss could have slowed the rotation of these  stars, this is inconsistent with their low mass-loss rates\footnote{We note that the absorption lines in HD~9974 have projected rotational velocity of 150-200 km s$^{-1}$ according to \citet{1981ApJ...244..173M}, arguing against a homogenous evolution explanation, unless the star is seen at an unfavorable inclination.  Is it an additional example of a WN3/O3?  The UV spectrum shows a ``weak" C\,{\sc iv} $\lambda 1550$ line according to \citet{2004MNRAS.353..153M}.  The \citet{2004MNRAS.353..153M} study of HD~9974 reported an absolute magnitude $M_V=-3.7$, twice as bright as our LMC WN3/O3s using their adopted 4.3~kpc kinematic distance, but \citet{2020MNRAS.493.1512R} find a Gaia distance of 2.9~kpc.  That would bring HD~9974's absolute visual magnitude $M_V$ to $-2.9$, similar to what we found for our WN3/O3s.   Since the mass-loss and other physical parameters were computed using a luminosity derived from an incorrect distance, we conclude that the analysis needs to be redone in order to answer the nature of HD~9974.}.

A third option is that these WN3/O3s are the products of binary evolution.  Although we can exclude the possibility that they have massive, luminous companions at present, we need to consider whether they have a lower-mass companion.  \citet{NeugentWN3O3s} argues that it is unlikely that such a putative companion is a neutron star or black hole, as none of the WN3/O3s is an X-ray source. The lack of X-rays rules out these being close binaries with compact companions.  However, a compact companion might remain bound but be in a wide orbit (and hence not generating X-rays) after it stripped off the outer layers of the WN3/O3 precursor. \citet{NeugentWN3O3s} note that such a situation would occur only through a narrow range of initial conditions, and the lifetime of the remaining star would be quite short as shown by \citet{1994A&A...290..119P}.   Another binary scenario would be for a main-sequence star to merge with an early-type WN star, enriching its surface with hydrogen, and forming the WN3/O3.  However,
\citet{NeugentWN3O3s} suggest that in such a scenario one would again expect the remaining object to be a rapid rotator, which---thanks to the absorption line profiles---we can rule out. 

Regardless of these arguments, we have long planned to carry out a radial velocity study of the WN3/O3s to either find companions to the WN3/O3s or set stringent limits on their existence.  In  \citet{FinalCensus} we gave a brief progress report, and confidently asserted that our study would be completed by the following Magellanic Cloud season. Unfortunately, weather-related delays and the global COVID-19 pandemic delayed our work until now.

In Section~\ref{Sec-obs} we describe our observations and reductions. In Section~\ref{Sec-RVs} we describe our procedure for measuring the radial velocities and present the data from our multi-year study. In Section~\ref{Sec-limits} we describe what these measurements mean in terms of setting limits on any companion, and in Section~\ref{Sec-sum} we summarize and discuss the implications of our findings\footnote{A tenth star, LMCe055-1, has properties somewhat similar to the WN3/O3s.  Classified as a WN4/O4, it is faint, and our preliminary modeling shows that the emission and {\it most} of the absorption arises in a single object.  A faint He\,{\sc i} $\lambda 4471$ feature, however, comes from a companion object, and the star eclipses.  We will present the results of our analysis of our comprehensive photometry and spectral modeling in a subsequent paper.}.

\clearpage

\section{Observations and Reductions}
\label{Sec-obs}

Although our initial plan was to obtain multiple spectra of each of the nine WN3/O3s, as observing  seasons came and went, we chose instead to concentrate on fewer stars but obtain more spectra. 
In the end, we obtained enough spectra for six of the WN3/O3s to adequately look for the presence of lower mass companions.  We list these stars in Table~\ref{tab:stars}.  We include basic information repeated from \citet{FinalCensus}, as well as listing the number of spectra we obtained.  As we show below, for measuring the radial velocities of the weak absorption lines adequately, we needed to use only the spectra with SNRs$>$100, and we quote that number as well.

Our discovery spectra were all taken with the Las Campanas Magellan Echellette (MagE) spectrograph \citep{MagE}, and we continued to use this instrument for our follow-up radial velocity measurements owing to its excellent throughput and good spectral resolution.  For the data taken in 2014 and 2015, the instrument was mounted on the Clay 6.5-meter Magellan telescope, after which the instrument was moved to the Baade 6.5-meter Magellan telescope.  The 1\arcsec\ wide slit was used, resulting in a spectral resolving power $R=4100$.   Wavelength coverage was from the atmospheric cutoff in the near-UV ($\sim$3200\AA) to 1$\mu$m.  Data were taken either as fill-in on other observing projects, or on dedicated one- or two-night 
observing runs.  The slit was oriented to the parallactic angle, except occasionally for LMC172-1, where we needed to keep a nearby star off the slit. (This companion is 2~mag brighter optically, and is located 3\farcs2 to the west.)  

The challenges of flat-fielding an echellette with such a wide wavelength range are severe, especially since our goal was to obtain spectra with signal-to-noise ratios (SNRs) of 100 or higher. \citet{2012ApJ...748...96M} found they could achieve SNRs $>$350 with MagE by
{\it not} flat-fielding their data, but rather by dithering along the 10\arcsec\ long slit.  We did not need SNRs
that high for this project, and so we did not dither, but instead relied upon the intrinsic uniformity of the CCD, following a suggestion by Ian Thompson \citep{MasseyHanson}. We did use well-exposed dome-flat exposures to
flat-field in the red to remove the fringing. Bias frames were also obtained and used to subtract from the data, although there is negligible bias structure.  After each set of observations of a star, a 3 sec long Th-Ar lamp exposure was made to provide wavelength calibration before moving the telescope to the next object. Several spectrophotometric standards were observed each night to provide flux calibration, useful for combining the spectral orders. Reductions were carried out using a combination of standard {\sc iraf} routines and special scripts written by Jack Baldwin.  Further reduction details can be found in \citet{2012ApJ...748...96M}.  

In Table~\ref{tab:exposures} we list the details of the observations of each star.  Our exposures ranged from
a single 10 min exposure to an hour (3$\times$20 min).  The observations with short exposure times were made before we began velocity monitoring; i.e., they were primarily our ``discovery" data, and  proved not to have sufficiently high SNRs to be used for our radial velocity study. They are included in the Journal of Observations only as they have been mentioned in earlier works.  
As discussed below,
we found that the cross-correlations were more reliable with spectra of SNR values (per 3-pixel spectral resolution element) of 100 or greater.  Those spectra are indicated with a letter designation in Table~\ref{tab:exposures} that will be used to identify the results of the radial velocity cross-correlation results in the next section.  A few spectra that are listed with poorer SNRs in Table~\ref{tab:exposures} were obtained with poor seeing and/or through clouds, both relatively rare occurrences on Las Campanas during Magellanic Cloud observing seasons.

As argued by \citet{FinalCensus}, our choice of MagE proved optimal for this project.  Residuals from the wavelength solutions were typically 0.05-0.06\AA\ (3 km s$^{-1}$ in the blue). Although better wavelength calibration could be achieved with a higher dispersion instrument, such as MIKE (the Magellan Inamori Kyocera Echelle), using it rather than MagE would have provided no greater accuracy.   The absorption lines are to 120-150 km s$^{-1}$ and thus well sampled with our MagE 3-pixel spectral resolution ($R=4100$, or 73 km s$^{-1}$). Even the most narrow emission line, N\,{\sc v}$\lambda$4946, has a width of 300 km s$^{-1}$. Thus the WN3/O3 spectra features are well sampled with MagE, and for a given exposure time we achieve a 2.5$\times$ larger SNR per spectral resolution element than we would have with MIKE.

\section{Radial Velocity Analysis}
\label{Sec-RVs}

\subsection{Methodology}

The difficulty of measuring stellar radial velocities is dependent on spectral type.  For most stars of type F and later, cross-correlation of a star's spectrum with a suitable radial velocity template is standard technique, utilizing many dozens of spectral features \citep{1979AJ.....84.1511T}, with precisions now reaching tens of cm~s$^{-1}$ employed to find extra-solar planets (see, e.g., \citealt{2018haex.bookE...4W,2022AJ....163..171Z}).  Main-sequence stars of earlier types have fewer, and broader, lines, and traditionally lines are measured one-by-one and the results averaged (see, e.g., \citealt{2004NewAR..48..727N, 2014ApJ...789..139M}) although sometimes also by cross-correlation techniques (e.g., \citealt{2008ApJ...682L.117G}).  

Determining the radial velocities of WR stars is particularly challenging, as the emission lines are all formed in an accelerating stellar wind.  This has two consequences: (a) lines of different ionization levels will have different velocities depending upon
their location in the wind, and (b) most lines are incredibly broad, with widths of thousands of km~s$^{-1}$, making precise measurements difficult.  A variety of techniques have been used over the years, including Gaussian fitting and intensity centroids (e.g., \citealt{2002MNRAS.333..347N}) and centroids of higher order (e.g., \citealt{1980ApJ...236..526M}).  

However, we are not so much interested in the actual radial velocity of our WN3/O3s.  Rather, our goal is to determine if our spectra show statistically significant radial velocity variations over time, and, if not, to place upper limits on any radial velocity variability.  This will allow us to place upper limits on the masses of any undetected companions.

A particular powerful technique for such work was pioneered by \citet{NeugentBinaries} to establish the relative binary frequency of WRs in M31 and M33.   They utilized a cross-correlation method using
selected spectral regions.  Cross-correlation techniques against a WR ``standard" is unlikely to work well, as line profiles usually differ significantly from star to star. Instead, \citet{NeugentBinaries} cross-correlated each spectrum of a given star against each of the other spectra of the same star.  Furthermore, since the density of spectral features is low compared to that of late-type ``normal" stars, they did this cross-correlation on selected spectral regions, isolating one or two features since the line-free continuum adds only noise. The cross-correlation produces a measure of the velocity shift between one spectrum and another for each of
these regions. The dispersion in these shifts from spectral region to spectral region for a particular cross-correlation pair tells us the measuring uncertainty.  A comparison of this ``internal error" ($I$) with the average velocity shift from pair to pair (the ``external error," $E$), corrected for the number of lines measured, forms the basis for concluding if the data indicate the star is a binary.  

As an example, consider the case where the dispersion in radial velocity measurements of ten spectral lines on each spectrum is, on average, 10 km s$^{-1}$ ($I$).   If the star is a binary with an orbital semi-amplitude $K=100$ km~s$^{-1}$ (i.e., a full amplitude of 200~km~s$^{-1}$), and one has 9 spectra taken at random orbital phases, then a Monte Carlo simulation shows that
the dispersion in the averages of the resulting 36 unique cross-correlation pairs (A-B, A-C, A-D,...B-C, B-D, B-E,...G-H, G-I, H-I) will be 70~km~s$^{-1}$.   To compare this to the internal error, we have to adjust the ratio by the square root of the number of lines that were measured in producing these errors, resulting in an $E/I$ value of 22.  We would reasonably conclude the star is a binary.  If 
instead $K=10$~km~s$^{-1}$, the average dispersion from pair to pair will be 7.5~km~s$^{-1}$, and $E/I$=2.6, again suggestive of binary motion, although if the measurement had been based upon fewer lines, we would be less convinced. With only four lines, we would obtain an $E/I$ of 2.1.    In other words, the larger the value of $E/I$, the more likely the pair-to-pair averages represent actual changes.  

Typically values of E/I greater than 2 are taken as evidence of statistically significant variations.  This rule-of-thumb was not based on statistics, but seems to trace back to an empirical determination by \citet{1969PASP...81..332A}.  (See also \citealt{1974AJ.....79.1307P}.) However, as
various authors have pointed out (see e.g.,  \citealt{1977ApJ...215..561C}), the actual statistical probability corresponding to a particular $E/I$ value depends upon both the number of lines and number of spectra.
This classic $E/I$ test is a simplified version of the general analysis of variance (ANOVA) test, which can be used to compute the F distribution, which takes into account the degrees of freedom, unlike the $E/I$ computation. This F distribution (or ratio) tests the hypothesis that there is no variability.
Values with probabilities less than 1\% likely mean that there are statistically significant velocity variations.  Using $E/I\geq2$ as a measure of binarity is statistically very conservative and likely to miss real binaries, as $E/I=2$ corresponds to probabilities of less than 0.5\% for six spectra with eight lines per spectrum, and less than 0.01\% for twelve spectra with eight lines per spectrum \citealt{1977ApJ...215..561C}.)\footnote{We note that by using the relative velocities, rather than the absolute velocities, we can determine
probabilities using the one-way ANOVA test.  \citet{1977ApJ...215..561C} and \citet{garmany} describe using two-way ANOVA computations in the case of spectral lines having formed in regions of differing outflows, a situation that they demonstrate is true for the absorption lines of O-type stars.  Sadly, these findings have typically been overlooked in recent analyses of the binary frequency of O-type stars.}

Consider three additional examples.  As noted above, a $K=10$ km s$^{-1}$ binary with an average dispersion of 10 km s$^{-1}$ from measuring four spectral lines on 9 spectra, our $E/I$ test would yield 2.1, just barely above the nominal cutoff.  An ANOVA test yields an F value of 2.9, corresponding to a probability of 0.001\%, well below our 1\% criteria: we would say that the data definitely supports the star being a binary.  A marginal case would be a $K=5$ km s$^{-1}$ binary with the same 10 km s$^{-1}$ measuring uncertainty and the same amount of data.  In that case,  F=1.62, corresponding to a probability of 3\%.  We would consider such data ``suggestive" but inconclusive.  A system where $K=1$ km s$^{-1}$  would certainly be undetected with a 10~km s$^{-1}$ measuring uncertainty: F=0.6, corresponding to a probability of 94\%, saying that any spectrum-to-spectrum variations are lost within the internal error. As our good friend and colleague, the late Virpi Niemela would say, one can never rule out any star as being a binary, but what we could say in such a case is that the data do not
support binarity.  However, our analysis would be able to place limits on any binary motion.  

Having described in detail the $E/I$ statistic and the ANOVA test, we will now apply these methodologies
to the measurements of our spectra.  This will allow us to determine if our data indicate binarity, and if not, to determine what the upper limits are on the masses of any undetected companions.

\subsection{Measurements and Results}

\begin{figure}
\epsscale{0.5}
\plotone{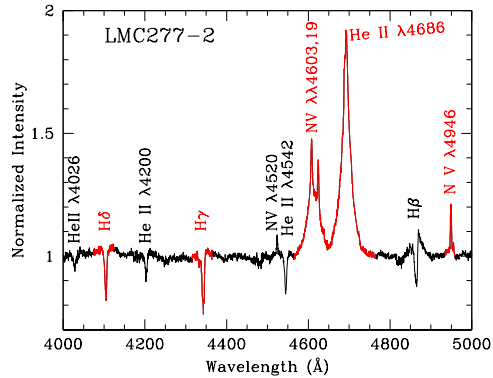}
\caption{ \label{fig:plotme}  A section of one of our highest SNR spectra.  The regions used in our
cross-correlations are shown in red.}
\end{figure}

In order to identify the wavelength regions that give us the most robust velocity measurements, we experimented with two spectra of LMC277-2 taken on the same night (2018 Jan 6) a few hours apart. In the end, we chose four wavelength regions for use in our cross-correlations for this project, two containing emission features and two containing absorption lines.   For the emission, we chose 4934-4959\AA, which contains the N\,{\sc v} $\lambda 4946$ line, and 4565-4765\AA, which contains the N\,{\sc v} $\lambda \lambda 4603,19$ doublet and the He\,{\sc ii} $\lambda 4686$ line.   The N\,{\sc v} $\lambda 4946$ feature is a particularly narrow,  strong line, while the N\,{\sc v}-He\,{\sc ii} complex contains the strongest emission features. For the absorption, we chose the 4075-4125\AA\ region, containing the H$\delta$/He\,{\sc ii} $\lambda$4100 blend, and the 4315-4365\AA\ region, containing the H$\gamma$/He\,{\sc ii} $\lambda$4339 blend. 
These are the strongest and best defined of the absorption features, as the higher-order Balmer/Pickering lines have poorer SNRs, while the H$\beta$/He\,{\sc ii} $\lambda$4859 and H$\alpha$/He\,{\sc ii} $\lambda$6560 have strong emission components.  The He\,{\sc ii} odd-n Pickering lines
(e.g., $\lambda$ 4200, 4542, 5411) were simply too weak and broad to provide reliable velocities.
We illustrate the four regions chosen in Figure~\ref{fig:plotme}\footnote{Note that although lines like the He\,{\sc ii} $\lambda 4200$ and $\lambda 4542$ stand out well in this very high SNR spectrum, they resulted in poorer results in the spectra with SNRs of only 100.}.

Before cross-correlation, the spectra were prepared by trimming (3785-6850\AA) in order to ease the normalization process.  The spectra were then normalized within {\sc iraf} using a fifth-order cubic spline with iterative rejection of 2$\sigma$ high and low points.  After division, the intensity of the normalized spectra were shifted by subtracting 1.0 to minimize continuum contribution to the cross-correlations.  The {\sc iraf} routine {\sc fxcor} was then used for the correlations, using a 21-point wide parabolic fit. 

The measurements are given in Table~\ref{tab:measurements}.  The values are given in terms of the 
relative velocity between pairs of spectra for each wavelength region, after account for the (very slight)
differences in heliocentric corrections.  The designations for each pair were defined in Table~\ref{tab:exposures}.  Thus the first entry row for star LMC079-1 shows the velocity of 
spectrum A (taken on 2017 Feb 08, at an HJD of 2457792.606) minus the velocity obtained from spectrum B (taken on 2017 Dec 31, at an HJD of 2458118.659) for all four regions, followed by
the mean of these four values, and the standard deviation of those four values, all in km~s$^{-1}$.
At the end of the measurements for each star we give the standard deviation $\sigma_{\rm pairs}$ for each of the four wavelength regions, their mean, as well as the standard deviation of the means.

In Table~\ref{tab:results} we list the $E/I$ values
for each of our stars, as well as the results of our ANOVA analysis.  The internal error $I$ was computed as the average of the standard deviations of the measurements (i.e., the average of the last column in Table~\ref{tab:measurements}), while the $E/I$ values are 2$\times$ the mean $\sigma_{\rm pair}$ (taken from Table~\ref{tab:measurements}) of the means divided by $I$.   The F-ratio and corresponding probabilities were computed using the Python {\sc statsmodels.stats.anova.anova\_1m} function, and confirmed with the {\sc bioinfokit.analys anova\_stat} routine.
 
\clearpage

\section{Limits on Binarity of the WN3/O3s}
\label{Sec-limits}

The analysis of our spectra given in Table~\ref{tab:results} does not show any evidence of radial velocity variations for any of the six WN3/O3s in our sample.  Of course, low-amplitude velocity variations could be hidden by our measuring uncertainties.  In this section we will consider what limits we can place on binarity for the stars in our sample.

The internal errors $I$ in Table~\ref{tab:results} are roughly 10-14 km s$^{-1}$.    We conducted multiple simulations to see what these meant in terms of what orbital semi-amplitude $K$ could be hidden in our data.  We find that with 28 cross-correlation pairs (corresponding to 8 spectra), and 4 measurements, a circular orbit with $K\ge10$ would be reliably detected as having statistically significant ($<$1\%) radial velocity variations as long as the period was shorter than the span of time over which our observations were made.  Thus we assume that if there were a companion in any of these systems, its orbital motion is less than this.

What does this mean in terms of a mass for such a hypothetical, unseen companion?  We will use $K=10$ km~s$^{-1}$ to compute the mass function $f$ as a function of period, and then use this to 
set limits on the mass of the companion star.  For this, we must know the mass of the WN3/O3 star.
Thanks to the presence of the absorption lines, \citet{NeugentWN3O3s} were able to determine surface gravities for the WN3/O3s; combined with the other physical properties they derived, these provided mass estimates good to 20\%.
Values for the stars in our sample ranged from 9$M_\odot$ (LMC199-1) to 19$M_\odot$ (LMC277-2).  Thus we will assume that the mass of the WN3/O3 component is 14$\pm5M_\odot$.

The mass function $f$ is related to the masses and orbital parameters as $$f=\frac{M_c^3 \sin^3i}{(M_{\rm WN3/O3}+M_c)^2}=\frac{P K^3}{2\pi G}(1-e^2)^{3/2},$$ where $M_c$ is the mass of the unseen companion, $P$ is the period, and $e$ is the orbital eccentricity. In solar units, and with $P$ in days and $K$ in km s$^{-1}$, $$\frac{M_c^3 \sin^3i}{(M_{\rm WN3/O3}+M_c)^2}=1.03\times10^{-7} P K^3 (1-e^2)^{3/2}.$$

\begin{figure}
\epsscale{0.37}
\plotone{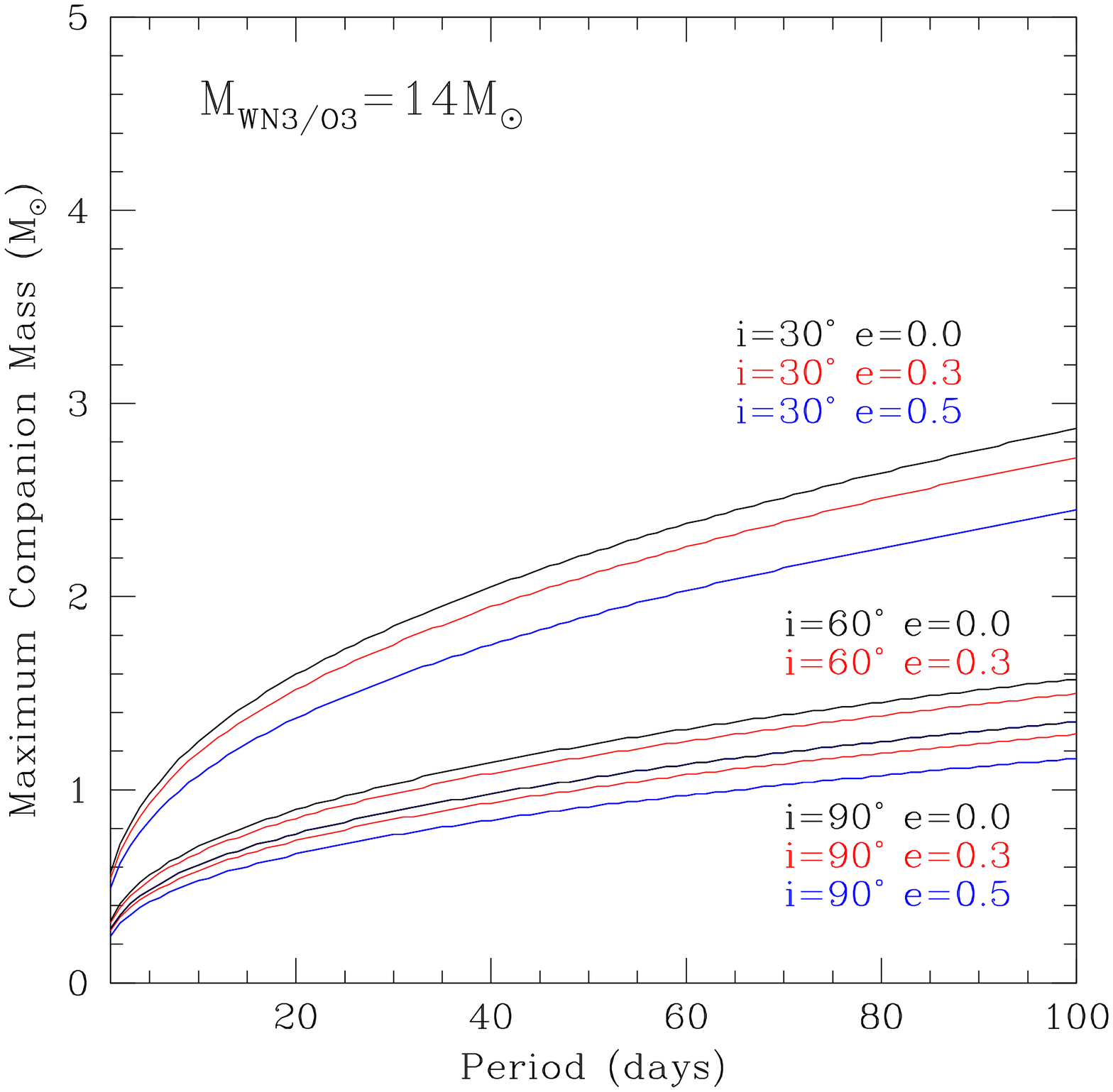}
\plotone{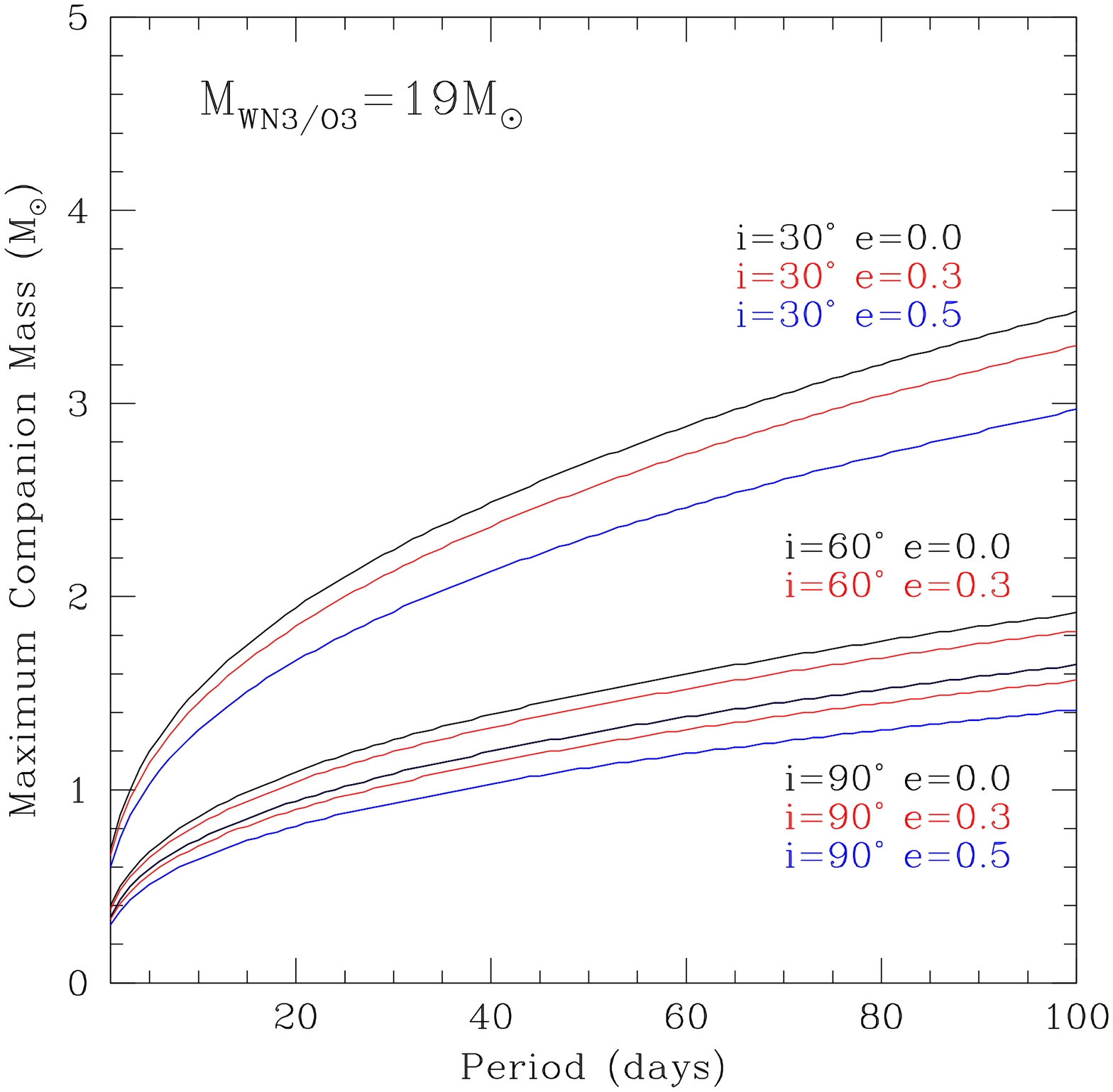}
\plotone{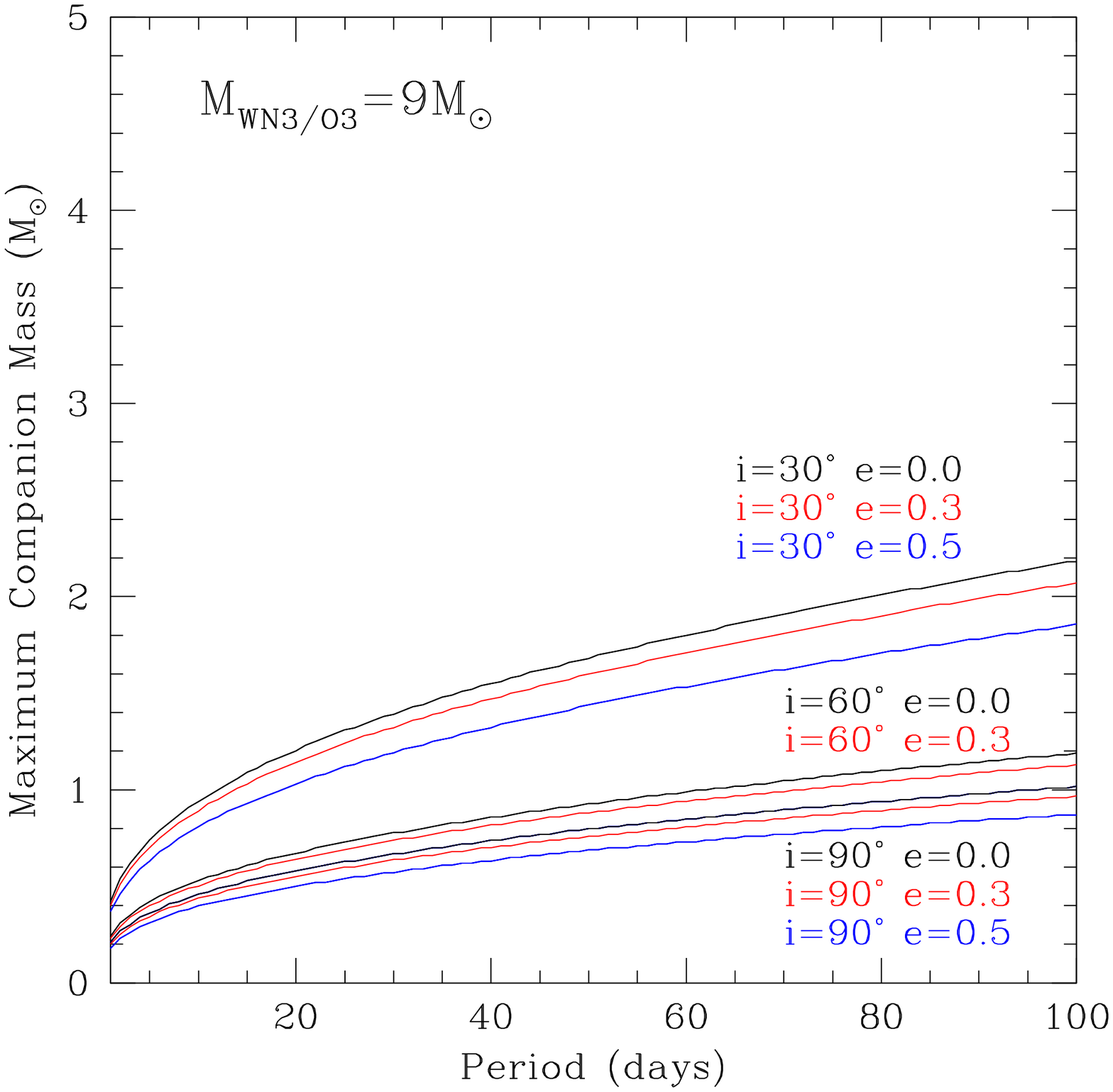}
\caption{{\label{fig:bme} The maximum allowable mass for any companion is shown as a function of period based upon the maximum orbital semi-amplitude allowed by our data (K$<$10 km~s$^{-1}$}).
The three panels cover the range of masses determined for the WN3/O3 stars in our sample by the analysis by \citet{NeugentWN3O3s}.  For each panel, we have computed 9 curves, corresponding to orbital inclinations $i$ of 90$^\circ$ (edge-on), 60$^\circ$, and 30$^\circ$, and eccentricities $e$ of 0.0 (circular orbit, shown in black), 0.3 (shown in red), and 0.5 (shown in blue). The $i=60^\circ$, $e=0.5$ curve is coincident with the $i=90^\circ$, $e=0.0$ curve.
}
\end{figure}

In Figure~\ref{fig:bme} we show the maximum mass of any companion star as a function of period,
based on the assumption that the largest orbital semi-amplitude $K$ that could be hidden in our data is  10 km s$^{-1}$.  For purposes
of illustration we include three orbital inclinations $i$, and three eccentricities $e$.  (Note that
if the orbit is elliptical, then our limit that $K<10$~ km s$^{-1}$ is not rigorous, but remains a good approximation.)  

We expect that the orientations of the orbital planes will be random.  The probability of an orbital inclination being between $i$ and $i+di$ will simply be proportional to the area subtended on a
unit sphere, 2$\pi\sin{i}\,di$; i.e., high inclination values are favored over low, and the expectation
value for the inclination $<i>$ is simply given by $$<i> = \frac{2\pi \int_{0}^{\pi/2}  i \sin{i} \,di} {2\pi \int_{0}^{\pi/2} \sin{i} \,di} = 1\ {\rm rad}.$$  This means that the 
$i=60^\circ$ (middle set) of curves in each panel is a good representation of the most probable
situation\footnote{This is nicely explained in http://keatonb.github.io/archivers/uniforminclination, which
also notes that $\cos{i}$ is uniformly distributed for isotropic inclination angles, a very useful result for Monte Carlo simulations.  See also Section 4.4 in \citet{2006asco.book.....H}.}.

Thus, any putative companion is likely to have a mass $\lesssim2M_\odot$ for periods less than 100 days.  Furthermore, a companion in a close orbit (period $<$10 days) would likely be solar or sub-solar in mass. We cannot rule out the possibility of extremely long periods (and thus higher masses) or unfavorable inclinations. 

\section{Summary and Discussion}
\label{Sec-sum}

We have conducted a radial velocity study of six of the nine known WN3/O3 stars, obtaining 6-8 high signal-to-noise spectra of each over a 3-5 yr period.  Our analysis shows no evidence of radial velocity variations. Any
binary motion would have to have an orbital semi-amplitude of $K\lesssim10$~km s$^{-1}$ to remain undetected in our data.  This requires that the mass of any unseen companion would likely be less
than 2$M_\odot$ for periods of 100 days or less, and less than 1$M_\odot$ for periods of 10 days or less.  Of course, in any individual case we cannot rule out the
possibility of an unfavorable inclination resulting in a higher mass companion going undetected, but that is unlikely to be the case for the entire sample.

The limits on the mass of a companion and the lack of X-ray emissions allow for the possibility of
a neutron star companion in a wide orbit.  Nor can we rule out the presence of a non-compact companion of solar mass.  However, much the same can be said of any WR star that lacks either X-ray emission and radial velocity variations.  We do note that the formation time for a solar-mass star is many times the age of a WR star, so such a hypothetical object would have to be in a T Tauri pre-main-sequence (PMS) stage. However, the initial mass $q$ ratio of such a WR plus 1$M_\odot$ PMS system would have to be $<$0.1.  No such systems are known to exist, and extrapolation suggests that such systems, if they were to exist, should be relatively rare \citep{2017ApJS..230...15M}.

Binary enthusiasts note that although only 40\% of WRs are found in massive binaries, the non-binary WRs may have been stripped by companions that has since merged.  However, as \citet{NeugentWN3O3s} argue, such a scenario should result in rapid rotation.  While this could be easily overlooked for most WRs, whose emission line widths are dominated by the stellar wind velocities, the WN3/O3s have absorption lines whose widths
give a good indication of the projected rotation rates, which are typical of normal O-type dwarfs, 120-150 km s$^{-1}$, as discussed earlier.  Also as discussed earlier, homogenous evolution also seems to be ruled out by the lack of rapid rotation.

Given that the binary fraction of WRs is 40\%, should we be concerned that {\it none} of the six WN3/O3s shows evidence of a companion?  We think not: the companions in most of the known WR systems are luminous O-type stars. Such a companion would dominate the spectral energy distribution, swamping intrinsic absorption from the WN3/O3 component.  Such a system might reveal itself to careful study: a WN3/O3+O9~V system,
for instance, would likely be classified as something like a WN3+O7~V, with the WN3/O3 contributing all of the He\,{\sc ii} absorption.  No obvious candidates are known in the LMC (see Table 3 in \citealt{FinalCensus}) or in the Milky Way (see \citealt{vanderHuchtII}), but perhaps the bright WN3+O7 SMC-AB6 or WN2+O6~V SMC-AB7 are examples of such composites\footnote{We also note that three other LMC stars have been identified by \citet{FinalCensus} as potential WN3/O3 candidates, BAT99 15a, Bat99 72, and BAT99 74.}.

In conclusion, our multiyear study fails to find evidence of binarity for any of the WN3/O3 stars.  We have presented arguments as to why past binarity is an unlikely explanation for their origins. That leaves us with the possibility that the WN3/O3s are a normal, short-lived transitional phase in the evolution of massive stars, where the stars have not shed all their hydrogen and their mass-loss rates are so low that their winds are still optically thin enough that we can see absorption lines.   Additional work is in progress, looking for other examples of this new class of WR in order to better understand their place in the evolution of massive stars.

\begin{acknowledgements}
Lowell Observatory sits at the base of mountains sacred to tribes throughout the region. We honor their past, present, and future generations, who have lived here for millennia and will forever call this place home.   The observations presented here were obtained over years at Las Campanas Observatory, and we are grateful to the excellent technical and logistical support we have always received there. We also acknowledge long-term support by both the Carnegie and Arizona Time Allocation Committees.
Partial support for this work was provided by the National Science Foundation through AST-83116 awarded to P.M.  In addition, support for K.F.N. was provided from NASA through the NASA Hubble Fellowship grant HST-HF2-51516 awarded by the Space Telescope Science Institute, which is operated by the Association of Universities for Research in Astronomy, Inc., for NASA, under contract NAS5-26555.  We are grateful to Drs.\ Michael Meyer and Trevor Dorn-Wallenstein for useful correspondence, and to an anonymous referee for suggestions that led to improvements in the paper.

\end{acknowledgements}

\clearpage

\begin{deluxetable}{l c c c c c r r}
\tabletypesize{\scriptsize}
\tablecaption{\label{tab:stars}WN3/O3 Stars in this Radial Velocity Study}
\tablewidth{0pt}
\tablehead{
\colhead{Star}
&\colhead{$\alpha_{2000}$}
&\colhead{$\delta_{2000}$}
&\colhead{$V$}
&\colhead{$B-V$}
&\colhead{$M_V$}
&\colhead{\#Obs}
&\colhead{\#SNR$>$100} 
}
\startdata
LMC079-1 & 05 07 13.33 & $-$70 33 33.9 & 16.31 & $-0.25$ & $-2.6$ &13 & 9 \\
LMC170-2 & 05 29 18.19 & $-$69 19 43.2 & 16.13 & $-0.17$ & $-2.8$  &10 & 6  \\
LMC172-1 & 05 35 00.90 & $-$69 21 20.2 & 15.95 & $-0.12$ & $-3.0$  & 12 &8 \\
LMC199-1 & 05 28 27.12 & $-$69 06 36.2 & 16.65 & $-0.22$ & $-2.3$  & 10 & 8 \\
LMC277-2 & 05 04 32.64 & $-$68 00 59.4 & 15.83 & $-0.16$ & $-3.1$  & 9 & 8 \\
LMCe159-1 & 05 24 56.87 &$-$66 26 44.4 & 16.34 & $-0.23$ &$-2.6$  & 9 & 8 \\
\enddata
\end{deluxetable}

\clearpage
\startlongtable
\begin{deluxetable}{l c r r c}
\tablecaption{\label{tab:exposures}Journal of Observations}
\tablewidth{0pt}
\tablehead{
\colhead{Date}
&\colhead{HJD}
&\colhead{Exp.\ time (s)}
&\colhead{SNR\tablenotemark{a}}
&\colhead{Designation} 
}
\startdata
\cutinhead{LMC079-1}
2013 Oct 18 & 2456583.869 & 1$\times$ 600 & 50 & \nodata \\
2013 Dec 14 & 2456640.671 & 1$\times$ 600 & 60 & \nodata\\
2015 Jan 09 & 2457031.668 & 1$\times$1200 & 80 & \nodata\\
2017 Feb 07 & 2457791.675 & 3$\times$ 500 & 60 & \nodata\\
2017 Feb 08 & 2457792.606 & 3$\times$ 550 & 100 & A \\
2017 Dec 31 & 2458118.659 & 3$\times$ 900 & 130 & B\\
2018 Jan 01 & 2458119.675 & 3$\times$ 900 &120 & C \\
2018 Jan 06 & 2458124.661 & 3$\times$ 900 &100 & D \\
2018 Feb 05 & 2458154.678 & 3$\times$ 900 &130 & E\\
2018 Nov 25 & 2458447.765 & 3$\times$ 900 & 140 & F \\
2020 Nov 26 & 2459179.684 & 3$\times$ 900 & 170 & G\\
2021 Dec 21 & 2459569.789 & 3$\times$ 900 & 130 & H\\
2022 Oct 02 & 2459854.738 & 3$\times$ 900 & 175 & I \\
\cutinhead{LMC277-2}
2013 Oct 18 & 2456583.844 & 1 $\times$ 600 &  75 & \nodata   \\
2013 Dec 14 & 2456640.662 & 1 $\times$ 600 &  70 & \nodata \\
2014 Sep 03 & 2457031.700 & 1 $\times$ 600 &  65 & \nodata \\
2015 Jan 09 & 2457031.670 & 3$\times$1200  & 120 & A  \\
2017 Feb 07 & 2457791.651 & 3$\times$ 600  &  65 & \nodata \\
2017 Feb 08 & 2457792.643 & 3$\times$ 600  & 105 & B    \\ 
2017 Dec 31 & 2458118.626 & 3$\times$ 900  & 120 & C  \\
2018 Jan 01 & 2458119.602 & 3$\times$ 900  & 125 & D \\
2018 Jan 06 & 2458124.696 & 3$\times$ 900  & 120 & E \\
2018 Nov 14 & 2458436.712 & 3$\times$ 900  & 145 & F  \\
\cutinhead{LMC172-1}
2013 Oct 16 & 2456581.723 &1$\times$ 600 & 75 & \nodata \\
2013 Dec 14 & 2456640.685 &1$\times$ 600 & 75 & \nodata \\
2015 Jan 09 & 2457031.813 &1$\times$1200 & 60 & \nodata \\
2017 Feb 07 & 2457791.626 &3$\times$600 & 95   & \nodata  \\
2017 Feb 08 & 2457792.667 &3$\times$ 600 & 120  & A \\
2018 Jan 06 & 2458124.730 &3$\times$ 900 & 120 & B \\
2018 Feb 04 & 2458153.668 &3$\times$ 900 &165 & C \\
2018 Nov 18 & 2458440.797 &3$\times$ 900&170 & D  \\
2018 Nov 26 & 2458448.631 &3$\times$ 900&130 & E\\
2020 Nov 26 & 2459179.638 &3$\times$ 900&195 & F\\
2022 Oct 02 & 2459854.790 &3$\times$ 900&190 & G \\
2022 Oct 31 & 2459883.823 &3$\times$1200&160 & H \\
\tablebreak
\cutinhead{199-1}
2013 Dec 14 & 2456640.637 & 1$\times$ 600 & 50 & \nodata \\
2015 Jan 09 & 2457031.829 & 1$\times$1500 & 80 & \nodata \\
2018 Jan 06 & 2458124.770 & 3$\times$1200 & 125 & A \\
2018 Feb 04 & 2458153.707 & 3$\times$1200 & 130 & B \\
2018 Nov 14 & 2458436.673 & 3$\times$1200 & 130 & C \\
2018 Nov 25 & 2458447.715 & 3$\times$1200 & 150 & D \\
2020 Jan 15 & 2458863.734 & 3$\times$ 900 & 120 & E \\
2020 Nov 25 & 2459178.750 & 3$\times$1200 & 180 & F \\
2020 Nov 26 & 2459179.589 & 3$\times$1200 & 165 & G \\
2020 Dec 09 & 2459192.699 & 3$\times$1200 & 130 & H \\
\cutinhead{277-2}
2013 Dec 14 & 2456640.662 &1$\times$ 600& 80 & \nodata \\
2015 Jan 09 & 2457031.686 &1$\times$1200&105 & A \\
2018 Jan 01 & 2458119.636 &3$\times$ 900&140 & B \\
2018 Jan 06 & 2458124.588 &3$\times$ 900&140 & C \\
2018 Jan 06 & 2458124.840 &3$\times$ 900&140 & D \\
2018 Nov 18 & 2458440.686 &3$\times$ 900&175 & E \\
2018 Nov 25 & 2458447.596 &3$\times$ 900&155 & F \\
2020 Jan 15 & 2458863.700 &3$\times$ 900&175 & G \\
2020 Dec 09 & 2459192.739 &3$\times$ 900&155 & H \\
\cutinhead{LMCe159-1}
2015 Jan 09 & 2457031.760 & 1$\times$1200 &  75 & \nodata \\
2018 Jan 06 & 2458124.626 & 3$\times$ 900 & 130 & A \\
2018 Feb 04 & 2458153.752 & 3$\times$1200 & 130 & B \\
2018 Nov 14 & 2458436.633 & 3$\times$ 900 & 125 & C \\
2018 Nov 18 & 2458440.831 & 3$\times$ 900 & 180 & D \\
2018 Nov 25 & 2458447.676 & 3$\times$ 900 & 165 & E \\
2018 Nov 26 & 2458448.665 & 3$\times$ 900 & 140 & F \\
2020 Jan 15 & 2458863.826 & 3$\times$ 900 & 160 & G \\
2020 Dec 09 & 2459192.773 & 3$\times$ 900 & 160 & H  \\  
\enddata
\tablenotetext{a}{Signal-to-noise ratio per 3-pixel spectral resolution
element measured over the region 4210-4330\AA.}
\end{deluxetable}

\clearpage
\startlongtable
\begin{deluxetable}{l r r r r r r}
\tablecaption{\label{tab:measurements}Radial Velocity Measurements}
\tablewidth{0pt}
\tablehead{
&\multicolumn{6}{c}{Radial Velocities (km s$^{-1}$)}\\ \cline{2-7}
\colhead{Cross}
&\colhead{N\,{\sc v}}
&\colhead{N\,{\sc v}+He\,{\sc ii}}
&\colhead{H$\delta$/He\,{\sc ii}}
&\colhead{H$\gamma$/He\,{\sc ii}}
&\colhead{Mean}
&\colhead{Std.\ Dev.}\\
\colhead{Pair\tablenotemark{a}}
&\colhead{$\lambda$4946}
&\colhead{$\lambda $4603-4686}
&\colhead{$\lambda$4100}
&\colhead{$\lambda$4339}
&\colhead{(regions)}
&\colhead{(regions)}
}
\startdata
\cutinhead{LMC079-1}
A-B    &    9.9 &    9.8 &   11.5 &   18.4 &   12.4 &    4.0 \\
A-C    &   12.6 &   11.1 &    4.6 &    8.6 &    9.2 &    3.5 \\
A-D    &   16.8 &    5.1 &   11.3 &   11.2 &   11.1 &    4.8 \\
A-E    &   13.0 &    1.5 &    0.3 &   14.4 &    7.3 &    7.4 \\
A-F    &   11.5 &  -13.8 &    1.1 &   11.8 &    2.7 &   12.1 \\
A-G    &    6.2 &    2.1 &    8.6 &    5.5 &    5.6 &    2.7 \\
A-H    &   16.5 &   19.6 &    2.6 &  -15.8 &    5.7 &   16.1 \\
A-I    &   13.8 &   -4.1 &   15.7 &   11.6 &    9.3 &    9.1 \\
B-C    &    2.9 &   -0.4 &   -8.5 &  -17.5 &   -5.9 &    9.1 \\
B-D    &    6.5 &   -3.4 &    0.8 &  -12.3 &   -2.1 &    7.9 \\
B-E    &    1.9 &   -7.7 &  -13.5 &   -6.9 &   -6.6 &    6.4 \\
B-F    &    1.0 &  -25.0 &   -6.2 &  -10.9 &  -10.3 &   11.0 \\
B-G    &   -2.3 &   -8.3 &   -2.4 &  -14.9 &   -7.0 &    6.0 \\
B-H    &    5.8 &    9.5 &   -8.8 &  -31.1 &   -6.1 &   18.4 \\
B-I    &    4.0 &  -14.3 &    7.7 &   -9.7 &   -3.1 &   10.6 \\
C-D    &    3.7 &   -4.4 &    7.9 &    0.6 &    2.0 &    5.2 \\
C-E    &   -0.7 &   -7.9 &   -6.3 &    1.7 &   -3.3 &    4.5 \\
C-F    &   -1.7 &  -22.2 &    2.1 &    1.9 &   -5.0 &   11.6 \\
C-G    &   -5.7 &   -7.9 &    4.5 &   -1.8 &   -2.7 &    5.4 \\
C-H    &    2.7 &    9.9 &    2.3 &  -27.0 &   -3.0 &   16.4 \\
C-I    &    1.3 &  -11.9 &   13.0 &    2.4 &    1.2 &   10.2 \\
D-E    &   -5.0 &   -4.2 &  -14.3 &    1.8 &   -5.4 &    6.6 \\
D-F    &   -4.0 &  -20.1 &   -7.7 &    4.4 &   -6.9 &   10.2 \\
D-G    &   -8.4 &   -3.3 &   -2.3 &   -3.1 &   -4.3 &    2.8 \\
D-H    &    0.4 &   14.4 &  -10.6 &  -21.0 &   -4.2 &   15.2 \\
D-I    &   -1.5 &   -9.6 &    4.5 &    2.5 &   -1.0 &    6.3 \\
E-F    &    0.2 &  -17.7 &    7.6 &   -0.1 &   -2.5 &   10.8 \\
E-G    &   -5.0 &    1.0 &   10.2 &   -9.7 &   -0.9 &    8.6 \\
E-H    &    5.3 &   16.1 &    4.8 &  -28.6 &   -0.6 &   19.4 \\
E-I    &    1.7 &   -6.5 &   17.6 &    0.3 &    3.3 &   10.2 \\
F-G    &   -4.0 &   15.5 &    2.7 &   -5.0 &    2.3 &    9.4 \\
F-H    &    4.7 &   31.1 &   -1.4 &  -26.5 &    2.0 &   23.6 \\
F-I    &    3.0 &   12.1 &   10.3 &    3.1 &    7.1 &    4.8 \\
G-H    &    8.9 &   15.7 &   -4.8 &  -20.3 &   -0.1 &   15.9 \\
G-I    &    7.5 &   -4.3 &    7.4 &    6.1 &    4.2 &    5.7 \\
H-I    &   -1.1 &  -19.4 &   16.5 &   24.2 &    5.1 &   19.5 \\
$\sigma_{\rm pairs}$&    6.4 &   13.2 &    8.4 &   13.8 &    5.7& \nodata \\
\tablebreak
\cutinhead{LMC277-2}
A-B    &  -13.6 &  -20.3 &    6.2 &   34.3 &    1.7 &   24.5 \\
A-C    &  -12.4 &   -8.4 &    4.2 &   34.3 &    4.4 &   21.1 \\
A-D    &  -10.8 &  -14.8 &   -2.9 &   33.2 &    1.2 &   21.9 \\
A-E    &  -13.5 &    0.3 &  -12.6 &   35.7 &    2.5 &   23.0 \\
A-F    &  -14.3 &  -25.5 &   -2.1 &   31.2 &   -2.7 &   24.5 \\
B-C    &    1.0 &   14.3 &    4.8 &    3.1 &    5.8 &    5.9 \\
B-D    &    4.2 &    6.1 &   -4.6 &    2.5 &    2.0 &    4.7 \\
B-E    &   -1.6 &   22.4 &  -12.3 &    4.7 &    3.3 &   14.5 \\
B-F    &    1.9 &   -5.0 &    1.4 &    1.5 &   -0.1 &    3.3 \\
C-D    &    2.4 &   -8.5 &    0.1 &   -0.1 &   -1.5 &    4.8 \\
C-E    &   -1.5 &    9.1 &  -11.6 &    0.0 &   -1.0 &    8.5 \\
C-F    &    0.8 &  -20.9 &    2.5 &   -3.5 &   -5.3 &   10.7 \\
D-E    &   -2.9 &   18.4 &   -7.2 &   -1.4 &    1.7 &   11.4 \\
D-F    &   -1.6 &  -13.5 &    1.7 &   -3.3 &   -4.2 &    6.6 \\
E-F    &    2.7 &  -28.8 &   17.8 &   -3.3 &   -2.9 &   19.4 \\
$\sigma_{\rm pairs}$&    6.9 &   16.2 &    8.1 &   16.7 &    3.2& \nodata \\
\tablebreak
\cutinhead{LMC172-1}
A-B    &    0.2 &   10.1 &   16.6 &   10.0 &    9.3 &    6.8 \\
A-C    &    4.3 &   13.0 &    4.4 &   22.9 &   11.2 &    8.8 \\
A-D    &   -3.7 &  -24.2 &   11.0 &    5.5 &   -2.8 &   15.5 \\
A-E    &   -5.3 &   -6.7 &    9.4 &    3.8 &    0.3 &    7.6 \\
A-F    &    1.7 &    1.8 &  -10.9 &   -8.2 &   -3.9 &    6.6 \\
A-G    &    6.3 &   -1.0 &    6.0 &   30.3 &   10.4 &   13.7 \\
A-H    &    1.6 &  -17.9 &   18.1 &   23.2 &    6.2 &   18.5 \\
B-C    &    3.0 &    3.5 &  -13.1 &    9.3 &    0.7 &    9.6 \\
B-D    &   -4.8 &  -32.0 &   -9.8 &   -5.4 &  -13.0 &   12.9 \\
B-E    &   -6.0 &  -16.7 &   10.6 &  -13.0 &   -6.3 &   12.1 \\
B-F    &    1.5 &   -7.1 &  -17.1 &  -28.3 &  -12.8 &   12.9 \\
B-G    &    6.2 &  -11.9 &    6.4 &   20.0 &    5.2 &   13.1 \\
B-H    &    2.8 &  -25.6 &   11.7 &    7.9 &   -0.8 &   16.9 \\
C-D    &   -9.3 &  -37.7 &    4.6 &  -22.2 &  -16.2 &   18.0 \\
C-E    &  -10.1 &  -20.3 &   10.8 &  -16.6 &   -9.1 &   13.9 \\
C-F    &   -2.1 &  -10.9 &   -9.3 &  -34.8 &  -14.3 &   14.2 \\
C-G    &    2.5 &  -16.4 &    7.5 &   13.3 &    1.7 &   12.9 \\
C-H    &   -0.6 &  -31.7 &   19.3 &   -0.9 &   -3.5 &   21.1 \\
D-E    &   -1.3 &   17.7 &    1.4 &    1.7 &    4.9 &    8.7 \\
D-F    &    6.7 &   27.5 &  -14.6 &   -4.6 &    3.8 &   18.1 \\
D-G    &   11.2 &   21.8 &   -9.7 &   34.0 &   14.3 &   18.5 \\
D-H    &    7.3 &    8.3 &   10.7 &   23.4 &   12.4 &    7.5 \\
E-F    &    8.0 &    9.2 &  -19.1 &  -14.4 &   -4.1 &   14.8 \\
E-G    &   12.2 &    4.2 &   -8.5 &   33.9 &   10.5 &   17.8 \\
E-H    &    7.5 &  -11.4 &    3.1 &   13.0 &    3.1 &   10.5 \\
F-G    &    4.3 &   -5.9 &    8.4 &   42.9 &   12.4 &   21.2 \\
F-H    &    0.4 &  -21.3 &   24.5 &   30.0 &    8.4 &   23.6 \\
G-H    &   -3.9 &  -15.2 &   17.2 &  -15.3 &   -4.3 &   15.3 \\
$\sigma_{\rm pairs}$&    5.7 &   17.0 &   12.2 &   20.4 &    8.8& \nodata \\
\tablebreak
\cutinhead{199-1}
A-B    &   -9.7 &   -8.7 &   18.4 &   -7.4 &   -1.9 &   13.5 \\
A-C    &   -4.2 &  -16.5 &   -8.8 &   11.0 &   -4.6 &   11.6 \\
A-D    &   -9.9 &  -23.3 &   -4.3 &  -15.8 &  -13.3 &    8.2 \\
A-E    &    7.2 &  -28.4 &   -7.3 &   19.2 &   -2.3 &   20.5 \\
A-F    &    1.5 &   -2.0 &   -8.4 &  -10.2 &   -4.8 &    5.5 \\
A-G    &   -0.4 &   -6.6 &   -7.5 &  -28.1 &  -10.6 &   12.0 \\
A-H    &   -5.8 &   -1.7 &   -5.1 &   -4.1 &   -4.1 &    1.8 \\
B-C    &    5.5 &   -9.0 &  -15.7 &   21.5 &    0.6 &   16.5 \\
B-D    &   -1.2 &  -14.3 &  -12.3 &   -0.4 &   -7.0 &    7.3 \\
B-E    &   16.7 &  -20.3 &  -10.0 &   26.5 &    3.2 &   22.0 \\
B-F    &   11.3 &    7.1 &  -13.8 &   -4.9 &   -0.1 &   11.5 \\
B-G    &    9.8 &    2.8 &  -22.3 &  -15.0 &   -6.2 &   15.0 \\
B-H    &    4.1 &    6.2 &  -15.1 &    4.9 &    0.0 &   10.1 \\
C-D    &   -7.4 &   -4.6 &    0.2 &  -20.5 &   -8.1 &    8.8 \\
C-E    &   11.2 &  -10.7 &    1.8 &   11.0 &    3.3 &   10.3 \\
C-F    &    5.0 &   14.9 &   -8.3 &  -17.1 &   -1.4 &   14.2 \\
C-G    &    4.9 &   11.3 &   -2.6 &  -31.3 &   -4.4 &   18.8 \\
C-H    &   -1.6 &   14.6 &   -1.0 &   -8.3 &    0.9 &    9.7 \\
D-E    &   18.9 &   -3.4 &    1.8 &   32.6 &   12.5 &   16.5 \\
D-F    &   12.9 &   21.0 &   -0.9 &    0.9 &    8.5 &   10.3 \\
D-G    &   12.3 &   15.4 &   -5.6 &  -11.4 &    2.7 &   13.2 \\
D-H    &    5.6 &   19.7 &   -4.0 &    8.6 &    7.5 &    9.8 \\
E-F    &   -5.5 &   27.5 &   -9.2 &  -52.5 &   -9.9 &   32.8 \\
E-G    &   -6.7 &   23.0 &  -14.5 &  -48.9 &  -11.8 &   29.6 \\
E-H    &  -12.7 &   25.3 &   -4.4 &  -20.6 &   -3.1 &   20.1 \\
F-G    &   -0.8 &   -4.0 &   -9.3 &  -14.5 &   -7.2 &    6.0 \\
F-H    &   -6.8 &   -0.2 &    9.4 &   15.1 &    4.4 &    9.7 \\
G-H    &   -5.6 &    2.9 &    3.6 &   21.6 &    5.6 &   11.5 \\
$\sigma_{\rm pairs}$&    8.6 &   15.2 &    8.2 &   21.2 &    6.4& \nodata \\
\tablebreak
\cutinhead{277-2}
A-B    &  -11.2 &  -17.7 &  -12.4 &   -1.8 &  -10.8 &    6.6 \\
A-C    &  -12.4 &  -10.5 &   -7.9 &  -12.4 &  -10.8 &    2.1 \\
A-D    &   -3.9 &  -20.1 &   -8.8 &    0.7 &   -8.0 &    8.9 \\
A-E    &   -5.2 &  -25.0 &   -6.0 &   -8.6 &  -11.2 &    9.3 \\
A-F    &   -5.5 &  -35.9 &  -10.6 &  -13.4 &  -16.4 &   13.4 \\
A-G    &  -18.0 &  -45.5 &   -3.0 &   15.3 &  -12.8 &   25.7 \\
A-H    &   -6.2 &  -19.5 &    6.7 &    0.4 &   -4.6 &   11.2 \\
B-C    &   -2.7 &    7.5 &    6.4 &   -8.3 &    0.7 &    7.5 \\
B-D    &    6.3 &   -1.3 &    4.9 &    3.8 &    3.4 &    3.3 \\
B-E    &    4.9 &   -9.0 &    2.5 &   -4.9 &   -1.6 &    6.5 \\
B-F    &    5.4 &  -22.7 &   -0.5 &   -8.7 &   -6.6 &   12.2 \\
B-G    &   -7.1 &  -27.3 &    7.4 &   21.4 &   -1.4 &   20.8 \\
B-H    &    3.5 &   -0.7 &   17.4 &    3.3 &    5.9 &    7.9 \\
C-D    &    8.5 &   -8.7 &   -1.3 &   12.2 &    2.7 &    9.5 \\
C-E    &    7.4 &  -16.8 &   -3.9 &    4.4 &   -2.2 &   10.8 \\
C-F    &    7.6 &  -28.9 &   -6.9 &   -0.1 &   -7.1 &   15.7 \\
C-G    &   -4.2 &  -36.3 &    1.0 &   29.8 &   -2.4 &   27.1 \\
C-H    &    6.2 &   -8.1 &   11.8 &   10.7 &    5.2 &    9.1 \\
D-E    &   -1.5 &   -7.7 &   -2.6 &   -7.9 &   -4.9 &    3.3 \\
D-F    &   -1.1 &  -21.2 &   -4.1 &  -12.1 &   -9.7 &    9.0 \\
D-G    &  -13.9 &  -27.3 &    1.8 &   18.0 &   -5.3 &   19.6 \\
D-H    &   -2.7 &    0.4 &   13.5 &   -0.6 &    2.6 &    7.4 \\
E-F    &   -0.1 &  -10.6 &   -3.0 &   -3.1 &   -4.2 &    4.5 \\
E-G    &  -11.7 &  -16.7 &    4.3 &   27.2 &    0.8 &   19.7 \\
E-H    &   -0.8 &    6.5 &   14.8 &    6.4 &    6.7 &    6.4 \\
F-G    &  -11.3 &   -6.4 &    7.9 &   31.5 &    5.4 &   19.2 \\
F-H    &   -0.8 &   20.1 &   18.1 &   10.6 &   12.0 &    9.5 \\
G-H    &   11.5 &   27.3 &    9.8 &  -19.1 &    7.4 &   19.3 \\
$\sigma_{\rm pairs}$&    7.7 &   16.5 &    8.5 &   13.5 &    7.1& \nodata \\
\tablebreak
\cutinhead{LMCe159-1}
A-B    &    1.2 &  -10.8 &    5.4 &   -3.5 &   -1.9 &    6.9 \\
A-C    &  -10.1 &  -12.5 &    5.9 &    2.8 &   -3.5 &    9.2 \\
A-D    &    5.6 &  -16.5 &   -5.8 &   -5.7 &   -5.6 &    9.0 \\
A-E    &    5.6 &    5.1 &  -10.1 &  -10.9 &   -2.6 &    9.2 \\
A-F    &   -6.3 &   -2.7 &   -7.8 &   -4.0 &   -5.2 &    2.3 \\
A-G    &    3.2 &   -1.6 &    0.0 &   19.1 &    5.2 &    9.5 \\
A-H    &    4.4 &    5.9 &    0.3 &   10.7 &    5.3 &    4.3 \\
B-C    &   -9.4 &   -2.7 &    2.5 &    5.9 &   -0.9 &    6.7 \\
B-D    &    5.0 &   -7.1 &   -6.7 &   -2.0 &   -2.7 &    5.6 \\
B-E    &    6.3 &   10.5 &  -16.2 &  -10.9 &   -2.6 &   13.0 \\
B-F    &   -6.0 &    5.7 &  -15.2 &   -5.3 &   -5.2 &    8.5 \\
B-G    &    2.7 &   11.1 &   -9.3 &   24.3 &    7.2 &   14.1 \\
B-H    &    4.0 &   18.0 &   -2.5 &   12.9 &    8.1 &    9.1 \\
C-D    &   15.1 &   -1.7 &  -10.5 &   -6.3 &   -0.9 &   11.2 \\
C-E    &   16.7 &   12.5 &  -20.7 &  -13.0 &   -1.1 &   18.5 \\
C-F    &    3.1 &    9.6 &  -15.8 &   -9.6 &   -3.2 &   11.6 \\
C-G    &   13.7 &   16.5 &   -7.2 &   26.0 &   12.3 &   14.0 \\
C-H    &   14.7 &   18.6 &   -8.7 &    7.1 &    7.9 &   12.1 \\
D-E    &    1.2 &   16.9 &   -5.4 &   -7.0 &    1.4 &   10.9 \\
D-F    &  -11.4 &   12.9 &   -5.5 &   -4.0 &   -2.0 &   10.4 \\
D-G    &   -1.8 &   18.2 &    2.8 &   28.0 &   11.8 &   13.8 \\
D-H    &   -0.8 &   22.0 &    3.0 &   16.6 &   10.2 &   10.9 \\
E-F    &  -13.4 &   -6.6 &   -1.7 &    5.4 &   -4.1 &    7.9 \\
E-G    &   -3.2 &   -2.0 &   10.2 &   35.6 &   10.1 &   18.0 \\
E-H    &   -1.6 &    2.6 &    9.2 &   21.4 &    7.9 &   10.0 \\
F-G    &   10.3 &    2.6 &   10.9 &   28.9 &   13.2 &   11.1 \\
F-H    &   11.9 &    7.8 &    8.3 &   15.9 &   11.0 &    3.8 \\
G-H    &    1.5 &    3.3 &   -0.6 &  -11.3 &   -1.8 &    6.5 \\
$\sigma_{\rm pairs}$&    8.2 &   10.3 &    8.5 &   14.6 &    6.3& \nodata \\
\enddata
\tablenotetext{a}{Identification of pair spectra are given by the ``Designation" column in Table~\ref{tab:exposures}.}
\end{deluxetable}

\clearpage
\begin{deluxetable}{l r r r r r r}
\tablecaption{\label{tab:results}Statistics of Radial Velocity Measurements}
\tablewidth{0pt}
\tablehead{
\colhead{Star}
&\colhead{$\sigma_{\rm pairs}$}
&\colhead{$I$}
&\colhead{$N$}
&\colhead{$E/I$}
&\colhead{$F$}
&\colhead{$p$} \\
& \colhead{km s$^{-1}$} 
& \colhead{km s$^{-1}$} 
}
\startdata
LMC079-1  & 5.72 &   9.76 & 8 & 1.17 & 1.07 & 0.39 \\
LMC170-2  & 3.20 & 13.66 & 6 & 0.47 & 0.17 & 1.00 \\
LMC172-1  & 8.82 & 13.96 & 8 & 1.26 & 1.43 & 0.11 \\
LMC199-1  & 6.39 & 13.45 & 8 & 0.95 & 0.72 & 0.83 \\
LMC277-2  & 7.07 & 11.63 & 8 & 1.22 & 1.11 & 0.34 \\
LMCe159-1 & 6.33 & 9.94 & 8  & 1.27 & 1.42 & 0.12 \\
\enddata
\end{deluxetable}

\bibliographystyle{aasjournal}
\bibliography{masterbib.bib}

\end{document}